\documentstyle[12pt]{article}

\begin{document}

\begin{titlepage}
December, 1992 \hfill                 UTHEP-250
\begin{center}

{\LARGE\bf The Spectrum of the Kazakov-Migdal Model}

\vspace{1.5cm}
\large{
S. Aoki\\
Institute of Physics, University of Tsukuba\\
Tsukuba, Ibaraki 305, Japan \\
\vspace{0.5cm}
and\\
}
\vspace{0.5cm}
\large{
Andreas Gocksch\\
Physics Department, Brookhaven National Laboratory\\
Upton, NY 11973, USA \\
}
\vspace{1.9cm}
\end{center}

\abstract{
Gross has found an exact expression for the density of eigenvalues
in the simplest version of the Kazakov-Migdal model of induced QCD. In this
paper we compute the spectrum of small fluctuations around Gross's
semi-circular
solution. By solving Migdal's wave equation we find a string-like spectrum
which, in four dimensions, corresponds to the infinite tower of mesons
in strong coupling lattice QCD with adjoint matter. In one dimension
our formula reproduces correctly the well known spectrum of the hermitean
matrix model with a harmonic oscillator potential. We comment on the
relevance of our results to the possibility of the model describing
extended objects in more than one dimension.
}
\vfill
\end{titlepage}

\section*{Introduction}
The original Kazakov-Migdal (KM) model of induced QCD \cite{KM}
consists of a scalar
in the adjoint representation of $SU(N)$, covariantly coupled to gauge fields
on a hypercubic lattice of spacing $a$. There is {\it no} kinetic term for the
gauge fields and the action of the model is
\begin{equation}
S = N \sum_x \left[ tr U(\Phi(x))-tr \sum_{\mu } \Phi(x)U_\mu(x)
\Phi(x+\mu a)U^\dagger_\mu(x)\right]~.  \label{km1}
\end{equation}
The hope was that this model could perhaps induce QCD in the sense
that a distances much larger than $a$ a kinetic term for the gauge fields
would be generated through scalar interactions at distances on the order
of the cutoff $a$. By carefully tuning the parameters of the potential
it might then be possible to reach an `asymptotic freedom domain' in which
the continuum limit could be taken. The mass of the scalar, which is kept heavy
in this limit, would act as an effective cutoff for the resulting
continuum (QCD?) theory.

The beauty of this idea of course lies in the fact that the model given
in Eq. (1) is analytically tractable in the limit of an infinite number
of colors $N$ \cite{migdal}.
The model is especially simple in the case of a purely
quadratic potential and initially there was clearly no a priori reason
for why the objective of inducing QCD, if possible at all, could not be
achieved by just tuning the mass of the scalar. This simplest of all
possibilities is by now ruled out. The reason is as follows: D. Gross
\cite{gross} has found an exact solution to the saddle point equations
describing the translationally invariant eigenvalue distribution of the
$\Phi$-field in the $N=\infty$ limit. This solution has {\it no} continuum
limit
in four dimensions. It rather describes the large $N$, infinite coupling
(remember, there is no plaquette term for the gauge fields in (1)) limit
of lattice QCD with adjoint matter. This in by itself is not sufficient to
prove `non-induction' for there could be other saddle points. However
computer simulations of the $SU(2)$ theory \cite{shen} also show that there
is no critical point in the case of a purely gaussian potential. Assuming
that the large $N$ limit is smooth the theorem is proved. Incidentally,
the simulations also showed that by adding a quartic term a critical point
could be reached. The nature of the resulting continuum theory is not clear
at the moment. Migdal \cite{migdal3} has made considerable progress in the
investigation of the theory with a general potential but lately he has
concentrated his efforts on the study of an extended version of Eq. (1),
the so called `mixed model' of induced QCD \cite{migdal2}.

In the mixed model \cite{migdal2} $n_f \ll N$ heavy fermions in the fundamental
representation
are added to the action in order to break the additional local symmetries
\cite{gross}
of the model, in particular the local $Z_N$ symmetry
$U_\mu(x) \rightarrow
Z_\mu(x)U_\mu(x)$
first discussed by Kogan et al. \cite{kogan}. As pointed out by these authors,
due to this symmetry the Wilson loop cannot acquire an expectation value. The
adjoint loop of course does, but it is always screened and cannot serve as an
order parameter for confinement. It is Migdal's hope that there is a phase
of the mixed model in which the center symmetry is broken and Wilson loops
show an area law.

In the present paper we will have nothing more to say about the question of
whether and how QCD can be induced. Rather we will extend  the very pretty work
of Ref. \cite{gross}
on the original Gaussian KM model by calculating the spectrum of small
fluctuations around the Gross saddle point. We believe that there are
many reasons for why this is interesting. In four dimensions our result
gives the masses of the mesons of the infinite coupling lattice gauge theory
to leading order in $ 1 / N$.
The large $N$, strong coupling spectrum of the theory
with fundamental fermionic matter has been known for over a decade now
\cite{smit}
and the adjoint scalar spectrum nicely complements this old work. Furthermore,
the model in Eq. (1) can be viewed as a gauged matrix model. Hermitean matrix
models in $D \leq 1$ dimensions
are intimately related to discretized versions of the Polyakov string \cite{KM}
and as such describe 2D gravity coupled to matter fields. In particular, a case
of great current interest is the one dimensional case, describing $c=1$ matter
coupled to 2D gravity. In more than one dimension ordinary matrix models
become very complicated \cite{kazakov} and one might hope that the gauged
version
in Eq. (1) might be useful as theory of extended objects in more than one
dimension.
It has also recently been suggested \cite{italians1} that the KM model can be
viewed in $D$
dimensions as the high temperature limit of the $(D+1)$ dimensional Wilson
action.
If this intersting
conjecture is indeed correct then our result in three dimensions
should describe the spectrum of fluctuations of the 4D-Polyakov line around
one of it's $Z_N$ minima at infinite temperature.
Finally, our calculation can serve as a testing ground for the methods
developed by Migdal. In some sense the present problem represents
the `hydrogen atom' of induced QCD; we can actually work out the eigenvalues
and the eigenfunctions of Migdal's wave equation in closed form. In the process
we also found a term in the wave equation which had been omitted in Ref.
\cite{migdal1}.

\section*{The Gross Solution}
In this section we will quickly review Gross's solution of induced QCD
with a quadratic potential. In the process we will also establish notation.
We will follow Migdal's approach \cite{migdal1} to the problem and express
the action in Eq. (1) in terms of the density of eigenvalues
\begin{equation}
\rho(\lambda)={1 \over N} \sum_i \delta(\lambda-\lambda_i).
\end{equation}
This is easily done by first integrating out the gauge fields using
the Itzykson-Zuber integral \cite{IZ}, which produces
\begin{equation}
S = N \sum_{x,i}  U(\lambda_i(x))- \sum_{x,i \neq j} \ln |\lambda_i(x)
-\lambda_j(x)|
-\sum_{x,\mu}\ln [I(\lambda(x),\lambda(x+\mu a)]
\end{equation}
where the logarithm comes from two factors of the Vandermond determinant
and $I$ denotes the Itzykson-Zuber determinant. Now using Eq. (2) the action
can be written in terms of a $x$-dependent density as
\begin{equation}
S[\rho] = N^2 \sum_{x} \int d \lambda \rho_x(\lambda) U(\lambda)
- N^2 \sum_{x} \int d\lambda \rho_x(\lambda) \int d\lambda^\prime
\rho_x(\lambda^\prime) \ln |\lambda -\lambda^\prime|
-\sum_{x,\mu}\ln [I(\rho_x,\rho_{x+\mu a})].
\end{equation}
Note that the effective action for $\rho$ will also receive contributions
from a Jacobian due to the change of variables from the eigenvalues
to the density \cite{migdal1}. Since everything we have to say is independent
of this term however we will just drop it here.

An equation for the eigenvalue distribution $\rho_x$ can be found by
looking for stationary points of the effective action in Eq. (4). In
performing infinitesimal variations of $\rho_x$ one must be careful though not
to change the normalization of $\rho_x$ in Eq. (2). This can be achieved by
taking variations of the form
\begin{equation}
\delta \rho_x (\lambda)= - {1 \over N} {d \psi(\lambda) \over d \lambda}
\end{equation}
or equivalently
\begin{equation}
{\delta \over \delta \psi(\lambda)} = {1 \over N} {d \over d \lambda}
{\delta \over \delta \rho(\lambda)}
\end{equation}
where $\psi(\lambda)$ vanishes at the end point of the support of
$\rho(\lambda)$.
The saddle point equation for (a translationally invariant) $\rho$
follows by setting
the first variation of the effective action
$ \displaystyle{\delta S \over \delta \psi(\lambda)}$
equal to zero. One obtains, denoting the logarithmic derivative of the
Itzykson-Zuber determinant by $F(\lambda)$,
\begin{equation}
F(\lambda)={-2 {\rm Re} V^\prime (\lambda) + U^\prime (\lambda) \over 2D}
\end{equation}
where
\begin{equation}
V^\prime (\lambda) = \int d\lambda^\prime {\rho (\lambda^\prime)
\over \lambda - \lambda^\prime}.
\end{equation}
This equation by itself does not
determine $\rho$ since $F$ is unknown. The function $F(\lambda)$ however
satisfies a set of Schwinger-Dyson equations which were derived by
Migdal in Ref. \cite{migdal}. Using these equations one finally ends up
with an integral equation for ${\rm Re} V^\prime (\lambda)$ whose imaginary
part determines $\rho$:
\begin{equation}
{\rm Re} V^\prime (\lambda)= P \int {d \lambda^\prime \over \pi}
\arctan{\pi \rho(\lambda^\prime )
\over \lambda - R(\lambda^\prime )}
\end{equation}
where
\begin{equation}
R(\lambda)= {D-1 \over D} {\rm Re} V^\prime (\lambda) + {U^\prime (\lambda)
\over 2D}.
\end{equation}

In one dimension the gauge field can be gauged away and
using $U(\Phi)={1 \over 2}m^2 \Phi^2$
the action in Eq. (1) is just that of a free scalar field.
In this case it is well known that $\rho$ is semicircular in shape, i.e.
\begin{equation}
\rho(\lambda)= \rho_0(\lambda)=
{1 \over \pi} \sqrt{\mu - {\mu^2 \lambda^2 \over 4}}  \label{semi}
\end{equation}
and $\mu = \sqrt{m^4-4}$.
Gross \cite{gross} realized that the semicircular form (\ref{semi})
for $\rho$ actually solves
Eqs. (9,10) in {\it any} number of dimensions. He showed that in D dimensions
\begin{equation}
\mu_{\pm}={m^2(D-1) \pm D \sqrt{m^4-4(2D-1)} \over 2D-1}
\end{equation}
and also computed the free energy\footnote{It is interesting to note that the
the free energy obtained by Kawamoto and Smit \cite{smit} can be brought into
the same
form if one makes the curious substitution $\lambda =
-\displaystyle{4 \over \mu^2}$ and
$M=\displaystyle{i m^2 \over 8}$,
where $\lambda$ is the square of the meson field and $M$
is the bare quark mass.}:
\begin{equation}
F={1 \over 2}{m^2 \over 2 \mu} + {1 \over 2}\ln \mu -{D \over 2} [
\sqrt{1+{4 \over \mu^2}} -1 - \ln ({1 \over 2} + {1 \over 2} \sqrt{1+{4 \over
\mu^2}})]
\end{equation}
The following things are important to note: $\mu_+$ is a minimum of $F$ and
vanishes at $m^2 =2D$ in $D\le 1$ implying that in this case the continuum
limit
can be taken. In $D >1$ on the other hand, $\mu_+$ never vanishes and instead
$\mu_- \to 0$ as $m^2 \to 2D^-$. However the saddle point at $\mu= \mu_-$ is
a local {\it maximum} which, as we shall see, has important consequences for
the
spectrum of the theory.

To summarize this section: For a quadratic potential the distribution
of eigenvalues is semicircular in shape in any number of dimensions.
In one or less dimensions a continuum limit can be constructed and we expect
a physical spectrum since one is perturbing around a local minimum of the free
energy.
In more than one dimensions on the other hand one expects tachyons in the
continuum limit. The spectrum of fluctuations around $\mu_+$ will describe
the spectrum of mesons made out of adjoint matter in the strong coupling
lattice theory. We will now go on to show all this explicitly.

\section*{Migdal's Wave Equation}
In order to get at the spectrum of the theory in leading order in
$\displaystyle {1 \over N}$
one must work out the effective action describing the fluctuations around
an extremum of the action (4). This has been done by Migdal \cite{migdal1}.
Writing
\begin{equation}
\delta \rho_x (\lambda)=\rho_x (\lambda )+
\delta \rho_x (\lambda)
\end{equation}
with $\delta \rho$ as in Eq. (5) and $\rho_x(\lambda)$ a solution of (7), one
obtains
\begin{equation}
S_2[\rho] =
-\sum_{x} \int d\lambda \int d\lambda^\prime \left[
{1 \over 2}\eta(\lambda,\lambda^\prime)\psi_x(\lambda) \psi_{x+a
\mu}(\lambda^\prime)
+[{1 \over (\lambda-\lambda^\prime)^2}+D \sigma(\lambda,\lambda^\prime)]
\psi_x(\lambda) \psi_{x}(\lambda^\prime)\right] .
\end{equation}
In the case at hand here we have $\rho_x (\lambda) =\rho_0(\lambda)$.
Now using plane waves, $\psi_x(\lambda)=\psi(\lambda) e^{iPx}$, one immediately
obtains Migdal's
wave equation for the particle spectrum:
\begin{equation}
\Omega^2\int d\lambda^\prime
\eta(\lambda,\lambda^\prime)\psi(\lambda^\prime)
=-\int d\lambda^\prime
[{1 \over (\lambda-\lambda^\prime)^2}+D \sigma(\lambda,\lambda^\prime)]
\psi(\lambda^\prime).
\end{equation}
In Eqs. (15,16) we have introduced Migdal's notation for the derivatives
of $F(\lambda)$, i.e.
\begin{equation}
\sigma(\lambda,\lambda^\prime)
= {d \over d \lambda^\prime}{\delta F(\lambda) \over \delta
\rho_x(\lambda^\prime)}
\end{equation}
and
\begin{equation}
\eta(\lambda,\lambda^\prime)
= {d \over d \lambda^\prime}{\delta F(\lambda) \over \delta \rho_{x+a
\mu}(\lambda^\prime)}.
\end{equation}
Also, in Eq. (16) we have defined $\Omega^2=\sum_{\mu}cos(P_\mu)$.
Migdal \cite{migdal1} has derived Integral equations for the functions $\eta$
and $\sigma$. We have checked his derivation of these equations and agree up
to an additional term in the equation for $\sigma$. The origin of this term
is explained in Appendix 1.
Using the correct equations for
$\eta$ and $\sigma$ from the Appendix we obtain, denoting
$\displaystyle{d \over d \lambda^\prime}\psi (\lambda^\prime) \equiv
H(\lambda^\prime)$ and integrating Eq. (16) over $\lambda$,
the following equation:
\begin{eqnarray}
\Omega^2\int d\lambda^\prime
{H(\lambda^\prime) \over \lambda^\prime -\lambda_0}=
\int d\lambda \int d\lambda^\prime H(\lambda^\prime) K(\lambda_0,\lambda)
\left[ {(1-D) \over \lambda -\lambda^\prime} +D {G(\lambda_0,\lambda^\prime)
\over G(\lambda_0,\lambda)} {1 \over \lambda - \lambda^\prime}\right] \nonumber
\\
-D \int d \lambda^\prime H(\lambda^\prime) G(\lambda_0,\lambda^\prime).
\label{wave}
\end{eqnarray}
Several comments are in order. First, all the integrals in Eq. (19) are defined
in the principal value sense. Second, Migdal's functions
$K(\lambda_0,\lambda)$ and $G(\lambda_0,\lambda)$ are given by
\begin{equation}
K(\lambda_0,\lambda)={\rho(\lambda) \over
\pi^2 \rho^2(\lambda)
+(\lambda_0 - R(\lambda))^2}
\end{equation}
and
\begin{equation}
G(\lambda_0,\lambda)= \displaystyle\frac{1}{\lambda_0 - R(\lambda )}
{\rm Re}\ \exp\left[+\int
{d \lambda^\prime \over \pi (\lambda^\prime-\lambda)}
\arctan{\pi \rho(\lambda^\prime )
\over \lambda_0 - R(\lambda^\prime )}\right].  \label{defg}
\end{equation}
Third, the range of the integrals is over the support of $\rho(\lambda)$ which
in our case is the interval $[-{2 \over \sqrt{\mu}}, +{2 \over \sqrt{\mu}}]$.
Fourth, the last term in Eq. (\ref{wave})
is the one absent in Migdal's paper \cite{migdal1}.

In principal the calculation of the spectrum of fluctuations around the Gross
saddle point is now straight forward: Compute $G(\lambda_0,\lambda)$ using it's
definition Eq. (\ref{defg}), do the $\lambda$-integral in (\ref{wave})
and finally solve the
resulting (singular) integral equation. However there are several subtleties
in the calculation which we thought make it worthwhile to present the
calculation
of $G(\lambda_0,\lambda)$ in some detail. This is done in Appendix 2. The
interested
reader will be able to check all the other integrals relevant here by using
the method used in the Appendix. Suffice it to say here, that boundary
conditions
{\it are} important. For example, the branch of the `$\arctan$' in
Eqs. (\ref{wave},\ref{defg}) is chosen
by a boundary condition at {\it infinity} \cite{migdal}. Hence, in the
calculation
$\lambda_0$ must be taken {\it outside} the support of $\rho_0$.

We have obtained the following expression for
$G(\lambda_0,\lambda)$:
\begin{equation}
G(\lambda_0,\lambda)={1 \over (R- {\mu \over 2})}\cdot
{(R \lambda_0-\lambda) -\sqrt{{\mu^2 \lambda_0^2 \over 4} -\mu} \over
(\lambda_0 - R \lambda)^2 + \mu -{\mu^2 \lambda^2 \over 4}}  \label{resg}
\end{equation}
where $R(\lambda)=R \lambda=\displaystyle{1 \over 2D}(m^2 +\mu (D-1))\lambda $
is linear in $\lambda$ for the semicircular
solution. To derive our final result for the wave equation below, we have
repeatedly used
the important identity
\begin{equation}
R^2 - {\mu^2 \over 4} =1
\end{equation}
satisfied by $\mu_\pm$.
Finally, for completeness we also give the results of the $\lambda$-integrals
in Eq. (\ref{wave}).
We obtained
\begin{equation}
\int d\lambda {K(\lambda_0,\lambda)
\over \lambda -\lambda^\prime}= {\mu^{3 \over 2} (R-xy) \over 4 \sqrt{x^2-1}
[(x-Ry)^2+{\mu^2 \over 4}(1-y^2)]}
\end{equation}
and
\begin{equation}
\int d\lambda K(\lambda_0,\lambda)
 {G(\lambda_0,\lambda^\prime)
\over G(\lambda_0,\lambda)} {1 \over \lambda - \lambda^\prime}
= {\mu \over 2}(1-{\sqrt{(Rx-{\mu \over 2}\sqrt{x^2-1})^2-1} \over
(Rx-y)-{\mu \over 2} \sqrt{x^2-1}})
\end{equation}
where we have introduced the new variables $x={\sqrt{\mu} \over 2}\lambda_0$
and $y={\sqrt{\mu} \over 2}\lambda^\prime$.

\section*{Calculation of Spectrum}
Using the results of the previous section we obtained the following
integral equation determining the spectrum of the theory to leading
order in $1/N$:
\begin{equation}
\Omega^2 \int_{-1}^{+1} dy{h(y) \over y-x} =
\int_{-1}^{+1} dyh(y) {A(x)-B(x)y \over (x-Ry)^2+{\mu^2 \over 4}(1-y^2)}
\end{equation}
where
\begin{equation}
A(x)={(1-D) \mu R \over 2 \sqrt{x^2-1}}-Dx
\end{equation}
and
\begin{equation}
B(x)={(1-D) \mu x \over 2 \sqrt{x^2-1}}-DR.
\end{equation}
To bring the wave equation into this form we have assumed that
$H(\lambda^\prime)$
depends on $\lambda^\prime$ only through $y$, i.e.
$H(\lambda^\prime)=h(y)$.

To determine the spectrum of the theory means to determine $\Omega^2$ in Eq.
(26).
Due to the compactness of the integration interval we expect a discrete
spectrum.
To solve the equation we follow the time honored method of expansion in a
complete,
orthonormal set of functions. In particular, due to the range of the
integration interval
and the form of the integrand on the left hand side of the equation, Chebyshev
polynomials are a natural set of functions to use here\footnote{We use the
conventions
of Ref. \cite{gradshteyn} for the definition of the Chebyshev polynomials.}.
Hence we make the ansatz
\begin{equation}
h(y)=\sum_n c_n {T_n(y) \over \sqrt{1-y^2}}.
\end{equation}
Using this ansatz the integral on the left hand side of Eq. (26) can be done
immediately
(remembering though that $|x| > 1$ in the calculation) and one obtains
\begin{equation}
\Omega^2 \int_{-1}^{+1} dy{h(y) \over y-x} =
-\pi \Omega^2 \sum_n c_n {(x-\sqrt{x^2-1})^n \over \sqrt{x^2-1}}.
\end{equation}
On the right hand side we first write
\begin{equation}
{A(x)-B(x)y \over (x-Ry)^2+{\mu^2 \over 4}(1-y^2)}={f(x) \over y-y_+} +{g(x)
\over y-y_-}
\end{equation}
where
\begin{equation}
y_{\pm}= Rx \pm {\mu \over 2} \sqrt{x^2 -1}. \label{defy}
\end{equation}
In the last equation the identity Eq. (23) has been used. Now the integral on
the
right hand side of the wave equation (26) can also be done and one obtains
after some
algebra,
\begin{eqnarray}
\int_{-1}^{+1} dyh(y) {A(x)-B(x)y \over (x-Ry)^2+{\mu^2 \over 4}(1-y^2)}
&=& \nonumber \\
-  {\pi \over 2}\sum_n c_n \{ [(D-1) +D] (R+{\mu \over 2})^{-n}
&+& [D-(D-1)](R+{\mu \over 2})^n\}
\displaystyle{(x-\sqrt{x^2-1})^n \over \sqrt{x^-1}}.
\end{eqnarray}
Now comparing Eqs. (30) and (33) we finally obtain for the eigenfunctions and
eigenvalues
\begin{equation}
h_n(y)={T_n(y) \over \sqrt{1-y^2}}
\end{equation}
and
\begin{equation}
{\Omega_n}^2={(D-1) \over 2}[ (R+{\mu \over 2})^{-n}-(R+{\mu \over 2})^n]
+{D \over 2}[ (R+{\mu \over 2})^n+(R+{\mu \over 2})^{-n}].  \label{res1}
\end{equation}
The last two equations constitute the main result of this paper. We will
discuss it's significance in the next and final section.
It is noted that the fluctuations themselves (at a given momentum $P_\mu$)
are of course given
by $  \delta \rho (y)= - \displaystyle {1\over N}
\displaystyle{T_n(y) \over \sqrt{1-y^2}}$ which
properly satisfies $\displaystyle \int_{-1}^1 dy \delta \rho (y) = 0$
for $n = 1,2,\cdots$, but $\not= 0$ for $n=0$. Therefore $h_0(y)$ should be
excluded from the eigenfunctions.

\section*{Discussion and Conclusion}
As expected we have found a discrete spectrum with an infinite number of
states.
The quantum number $n$ labels states according to their parity under
$y \rightarrow -y$. The eigenfunctions $h_n(y)$ are even (odd) under this
operation
for $n$ even (odd).
Let us begin our considerations of the spectrum now by
looking at the one dimensional case in some detail. To this end it is useful
to define
\begin{equation}
M=\ln (R+{\mu \over 2})
\end{equation}
and write Eq. (\ref{res1}) as
\begin{equation}
{\Omega_n}^2=D \cosh(nM) -(D-1) \sinh(nM) \label{spec}
\end{equation}
In one dimension we obtain
\begin{equation}
\cos(P_n)=\cosh(n\cdot{\rm arccosh}({m^2 \over 2}))=T_n({m^2 \over 2})
\end{equation}
The interpretation of this result is straightforward. As we mentioned before,
in one dimension the gauge fields play no role and the theory is free
with propagator
\begin{equation}
G_{ij,kl}(P)={1 \over N}{1 \over m^2 -2 \cos(P)}\delta_{il} \delta_{jk}.
\end{equation}
Clearly, the `pole' of the propagator is at $cos(P)={m^2 \over 2}$ which agrees
with (38) in the case $n=1$.
In position space this corresponds to an exponential falloff of the propagator
with mass $M$.
The states with $n >1$ correspond to `mesons' made
out of $n$ $\Phi$'s. The $x$-space propagator for these objects
decays exponentially with mass $n \cdot M$ which in momentum space gives
Eq. (38).

In one dimension $\mu=\sqrt{m^4-4}$ vanishes as $m^2 \rightarrow 2^+$
and we can take the continuum limit. Writing $m^2=2+m_0^2a^2$
and $P=P_0a$
one obtains
\begin{equation}
-P_0^2 \equiv E_n^2=n^2\cdot m_0^2.
\end{equation}
Hence in the continuum limit we get a linearly rising spectrum of states
of mass $n m_0$. Note, that for $m_0=0$ we get a massless boson in one
dimension
as opposed to the massless boson in $1+1$ (one spacelike, one timelike)
obtained
in the double scaling limit of the continuum model \cite{das}.
In less than one dimension $\mu$ still vanishes at $m^2=2D$ and one obtains
expanding Eq. (\ref{spec}) at $m^2=2D+m_0^2a^2$
\begin{equation}
E_n^2= n\cdot m_0^2
\end{equation}
in the continuum limit. Contrary to the case in one dimension, the particle
masses are proportional to $\sqrt{n}$.

Let us now go on and discuss the case $D >1$. As we pointed out before
the stable solution is $\mu= \mu_+$ in Eq. (12) which is always greater
than zero. In this case the interpretation of the spectrum in Eq. (\ref{spec})
in four dimensions is that of the mesons in strong coupling lattice QCD with
adjoint
matter.
Note that asymptotically in $n$, $ \Omega_n^2=(R+\mu / 2)^n$. The same formula
holds in the limit $D\rightarrow \infty$ for all $n$. It is interesting
to see what happens if instead we use $\mu=\mu_-$.
Setting $m^2=2D-m_0^2a^2$ in order to approach the critical point from below,
one gets
\begin{equation}
-P_0^2 \equiv E_n^2= -n\cdot m_0^2
\end{equation}
in the continuum limit.
Although infinitely many states appear in the continuum limit, they are all
$\it tachyonic$ as expected.  The spectrum in $D > 1$
could be improved if an upside-down
quadratic potential\cite{lat92} is used for $U(\Phi)$ in Eq. (\ref{km1}).
The continuum limit in this case is equivalent to setting $m^2=-2D-m_0^2a^2$
in Eq. (\ref{spec}), which gives
\begin{equation}
 E_n^2 a^2= 2D\ ( (-1)^n -1)+ (-1)^n \cdot n \cdot m_0^2a^2  .
\end{equation}
For $n$ even, we obtain $E_n = n\cdot m_0^2 > 0$ though
$E_n = -\infty$ for $n$ odd.

What do the above results tell us? In one dimension we got the expected
result. In more than one dimension we find an infinite tower of states
which certainly suggests a `stringy' interpretation of the spectrum.
Makeenko \cite{makeenko} has written down Schwinger-Dyson equations
for the functions
\begin{equation}
G(C_{xy})=< {1 \over N} tr(\Phi(x)U(C_{xy})\Phi(y)U(C_{yx}))>
\end{equation}
where $U(C_{xy})$ denotes a string of links along the path $C_{xy}$.
For the quadratic potential Makeenko has obtained the solution for $G(C_{xy})$
and it would be interesting to see whether by summing his result over all
paths our spectrum can be reproduced \cite{migdal2}\footnote{In one
dimension one just obtains
the square of the x-space propagator for a free particle by this procedure, i.e
one
only obtains the lightest `meson'.}.
If it is an interpretation
of our result in terms of the excitations of a string of flux seems natural.

\section*{Acknowledgements}
We thank V. Kazakov and A. Migdal for valuable discussions.
This work was supported in part by a DOE grant at
Brookhaven National Laboratory (DE-AC02-76CH00016).

\section*{Appendix 1. Derivation of the extra term in eq.(\ref{wave})}

In this Appendix we show how the last term in eq.(\ref{wave}) is derived.
In the Appendix of Ref. \cite{migdal1}
Migdal has derived the following equation (78)
\begin{equation}
\displaystyle\int d\mu
\displaystyle\frac{\rho (\mu ) g_\lambda (\phi' ,\mu )}{z - \mu }
= {\cal T}_\lambda (z) \displaystyle\int d\phi
\displaystyle\frac{\rho (\phi)}{z - \phi }
\displaystyle\frac{ (\sigma (\phi' ,\phi ) +
\frac{1}{G_\lambda (\phi )}\frac{d}{d\phi' }
\frac{G_\lambda (\phi' )-G_\lambda (\phi )}{(\phi'-\phi )} )}
{(\lambda - R(\phi ))^2 + \pi^2 \rho^2 (\phi) }
\end{equation}
where
\begin{equation}
g_\lambda (\phi' ,\phi ) =\displaystyle\frac{d}{d\phi'}
\displaystyle\frac{\delta G_\lambda (\phi )}{\delta \rho_x(\phi')} \ .
\end{equation}
Setting $z\rightarrow\infty$ in the above formula, we find
\begin{equation}
\displaystyle\int d\mu
\rho (\mu ) g_\lambda (\phi' ,\mu )
= \displaystyle\int d\phi
\rho (\phi)
\displaystyle\frac{ (\sigma (\phi' ,\phi ) +
\frac{1}{G_\lambda (\phi )}\frac{d}{d\phi' }
\frac{G_\lambda (\phi' )-G_\lambda (\phi )}{(\phi'-\phi )} )}
{(\lambda - R(\phi ))^2 + \pi^2 \rho^2 (\phi) }  \ .   \label{eq1}
\end{equation}
\  From eq.(59) in Ref. \cite{migdal1}
\begin{equation}
\displaystyle\int d\mu
\rho_x (\mu ) G_\lambda (\mu )
= \displaystyle\int d\nu \displaystyle\frac{\rho_{x+a\mu} (\nu )}{\lambda -\nu
} \ ,
\end{equation}
we obtain
\begin{equation}
\displaystyle\int d\mu
\rho (\mu ) g_\lambda (\phi' ,\mu )
= -\displaystyle\frac{d }{d\phi'}G_\lambda(\phi' ) \ ,  \label{extra}
\end{equation}
which differs from eq.(77) in Ref. \cite{migdal1}.
Inserting (\ref{extra}) into (\ref{eq1}), we finally find equation for
$\sigma (\phi' ,\phi)$
\begin{equation}
0
=\displaystyle\frac{d }{d\phi'}G_\lambda(\phi' ) + \displaystyle\int d\phi
\rho (\phi)
\displaystyle\frac{ (\sigma (\phi' ,\phi ) +
\frac{1}{G_\lambda (\phi )}\frac{d}{d\phi' }
\frac{G_\lambda (\phi' )-G_\lambda (\phi )}{(\phi'-\phi )} )}
{(\lambda - R(\phi ))^2 + \pi^2 \rho^2 (\phi) }  \ .
\end{equation}

\section*{Appendix 2. Calculation of $G(\lambda_0,\lambda )$}

The definition of $G(\lambda_0,\lambda )$ is
\begin{equation}
G(\lambda_0,\lambda)= \displaystyle\frac{1}{\lambda_0 - R(\lambda )}
{\rm Re} \left[ \ \exp [ \displaystyle\int { d \lambda^\prime \over \pi
(\lambda^\prime-\ \lambda)}
\arctan{\pi \rho(\lambda^\prime )
\over \lambda_0 - R(\lambda^\prime )} ] \ \right]   \label{g1}
\end{equation}
where $R(\lambda) = R \lambda$ for the semicircular solution
$\pi\rho(\lambda)= \sqrt{\mu - \mu^2 \lambda^2 / 4 }$.
We first calculate the following integral
\begin{equation}
J \equiv \displaystyle\int {+ d \lambda^\prime \over \pi (\lambda^\prime-\
\lambda)} \arctan{\sqrt{\mu - \mu^2 \lambda^2 / 4 }
\over \lambda_0 - R\lambda^\prime }   \ .
\end{equation}
Integration by parts together with the obvious rescaling of variables
gives
\begin{equation}
J = {\mu x\over\pi}\displaystyle\int_{-1}^1 {dy\over \sqrt{1-y^2}}\ln (y-z)
\ [ {f_+\over y-y_+}+{f_- \over y-y_-}]  \label{int1}
\end{equation}
where
$x={\sqrt{\mu}\over 2}\lambda_0$, $z={\sqrt{\mu}\over 2}\lambda$,
$y={\sqrt{\mu}\over 2}\lambda'$,
$
f_\pm= \pm \displaystyle {y_\pm - R / x\over y_+-y_-}\ ,
$
and $y_{\pm}$ are given in Eq. (\ref{defy}) in the text.
Using the following properties for Chebyshev polynomials $T_n(y)$
\begin{equation}
\ln (y-z) = - T_0(y)\ \ln (2t_z) - 2\sum_{n=1}^\infty T_n(y)\ {(t_z)^n\over n}
\end{equation}
\begin{equation}
{1\over y -y_\pm}= -{2t_{\pm}\over 1-t_\pm^2}\ ( \ T_0(y) + 2\sum_{n=1}^\infty
T_n\  (t_\pm)^n \ )
\end{equation}
\begin{equation}
\displaystyle\int_{-1}^1{dy\over \sqrt{1-y^2}}\ T_n(y)\cdot T_m(y)
= \delta_{nm} \times \left\{ \begin{array}{ll} \pi/ 2    & n\not= 0 \\
				    \pi          & n=0
		   \end{array}
           \right.
\end{equation}
where $t_z = z -i\sqrt{1-z^2}$ and $t_{\pm} = (R\mp {\mu\over 2})
(x-\sqrt{x^2-1})$,
we perform the integral in Eq. (\ref{int1}) and obtain
\begin{equation}
J = -\ln\ [\ {1-t_zt_+\over 1-t_zt_-} \ ]  \ .
\end{equation}
\ From the above result $G$ becomes
\begin{equation}
G(\lambda_0,\lambda) = {1\over \lambda_0 - R\lambda} {\rm Re}
\ [ \ {1-t_zt_+\over 1-t_zt_-}\ ]  \ ,
\end{equation}
which finally gives eq.(\ref{resg}) in the text.


\end{document}